\documentclass[aps,pre,twocolumn,superscriptaddress,showpacs,showkeys]{revtex4-2}

\usepackage{amsmath,amssymb}
\usepackage{graphicx,epstopdf}
\usepackage[normalem]{ulem}
\usepackage{color}
\usepackage{comment}
\usepackage{enumitem}
\usepackage[dvipsnames]{xcolor}
\usepackage[utf8]{inputenc}
\usepackage{float}
\usepackage{mciteplus}
\usepackage{chngcntr}
\usepackage{booktabs}
\usepackage{bigints}
\usepackage{relsize}
\usepackage[english]{babel}
\renewcommand\harvardurl[1]{\textbf{URL}: \url{#1}} 
\usepackage{xurl} 
\usepackage{hyperref}

\usepackage{ragged2e}
\usepackage{array,tabularx}
\usepackage{makecell}
\usepackage{soul}
\usepackage{lineno}

\newcolumntype{P}[1]{>{\centering\arraybackslash}p{#1}}

\definecolor{redcolor}{rgb}{1,0,0}
\def\Red#1{{\color{redcolor} #1}}

\definecolor{bluecolor}{rgb}{0,0,1}

\definecolor{pinkcolor}{rgb}{0.85, 0.45, 0.6}

\begin{document}

 \title{ Variable order porous media equations: Application on modeling the S\&P500 and Bitcoin price return} 
 	
\author{Yaoyue Tang \P } 
	\affiliation{School of Civil Engineering, The University of Sydney, Sydney NSW 2006, Australia}

\author{Fatemeh Gharari \P} 
	\affiliation{Department of Statistics and Computer Science, University of Mohaghegh Ardabili,  Ardabil, Iran}
 
 \author{Karina Arias-Calluari}
        \affiliation{School of Mathematics and Statistics, The University of Sydney, Sydney, NSW 2006, Australia}
    
 \author{Fernando Alonso-Marroquin}
	\affiliation{School of Civil Engineering, The University of Sydney, Sydney NSW 2006, Australia}
 \affiliation{CPG, King Fahd University of Petroleum and Minerals, Dhahran 31261,
Kingdom of Saudi Arabia}
	\email{fernando.alonso@sydney.edu.au}

 \author{M. N. Najafi}
   \affiliation{Department of Physics, University of Mohaghegh Ardabili, P.O. Box 179, Ardabil, Iran}
	\email{morteza.nattagh@gmail.com}

 \begin{abstract}
This article reveals a specific category of solutions for the $1+1$ Variable Order (VO) nonlinear fractional Fokker-Planck equations. These solutions are formulated using VO $q$-Gaussian functions, granting them significant versatility in their application to various real-world systems, such as financial economy areas spanning from conventional stock markets to cryptocurrencies. The VO $q$-Gaussian functions provide a more robust expression for the distribution function of price returns in real-world systems. Additionally, we analyzed the temporal evolution of the anomalous characteristic exponents derived from our study, which are associated with the long-range memory in time series data and autocorrelation patterns.

\end{abstract}

\maketitle


\section{Introduction }
Anomalous diffusion has manifested itself in various fields of science, such as  physics~\cite{solomon1993observation,chechkin2005fractional,umarov2009variable}, chemistry~\cite{magin2008anomalous,metzler2014anomalous}, biology~\cite{santamaria2006anomalous,guigas2008sampling}, and socioeconomic systems such as stock markets~\cite{alonso2019q,gharari2021space}. Although it was proposed for transport and wave propagation paradigms~\cite{richardson1926atmospheric,de1976percolation,west2001ant}, now its relation with other phenomena is well-established, including but not limited to fractals and percolation in porous media~\cite{puech1983fractal,mardoukhi2015geometry}, cell nucleus, plasma membrane and cytoplasm in biology~\cite{saxton2007biological}.
Anomalous diffusion is manifested in the process where the total displacement of the random walker scales with time exhibiting a fractional exponent, as a result of the correlations in the stochastic process~\cite{sokolov2005diffusion,pkekalski1999anomalous}.
Under specific assumptions, a fractional version of the Fokker-Plank equation (FPE) can be employed to describe the time evolution of these systems' probability density function (PDF). Non-local fractional derivatives are relevant when dealing with the Levy process, for example~\cite{yanovsky2000levy,anderson2018anomalous}. 
Intuitively, a non-local operator requires information from a whole interval when operating on a function, in contrast to local operators that only need information from a single point in their immediate vicinity \cite{almeida2016remark,gharari2021space}, for a comprehensive review, see Ref.~\cite{teodoro2019review} and the references therein. This fractionalization can occur in the \textit{time} and the \textit{space} derivatives of FPE. 
 While the linear fractional FPE is suitable for describing a wide range of systems with anomalous diffusion, its non-linear version has been implemented in more diverse domains, including biological systems~\cite{Murray1993epidemic}, thermostatistics~\cite{umarov2007multivariate,quiros1999asymptotic,gilding1976class,pamuk2005solution,plastino2020nonlinear}, and stock markets~\cite{ gell2004nonextensive, alonso2019q,gharari2021space}.  It was also shown that the PDF of the detrended price return of the S\&P500 index is governed by the porous media equation (PME)--- which is a non-linear FPE--- through a curve fitting analysis of the PDFs after collapsing self-similar $q$-Gaussian functions ~\cite{alonso2019q,alma991032099996005106,gharari2021space}. 
 
 Despite the relative success of the $q$-Gaussian distributions in explaining the time evolution of the PDF of many stochastic systems, some studies show that the numerically estimated exponents exhibit slow time dependence. The stock market is an example where the PDF of price return does not follow a constant order (CO) non-linear FPE, or at least has a limited validity~\cite{frame1970lassa}, as the price returns of stock market indexes exhibit characteristic exponents that depend on time~\cite{arias2021methods}, aligned with the central limit theorem (CLT). More specifically,  the stochastic fluctuations in the price return of the S\&P500 index can be modeled using superdiffusive self-similar $q$-Gaussian functions and the anomalous diffusion exponent $\alpha$, which has initial values ($\alpha>2$, $q>1$), and then slowly converge to  $\alpha\to 2,q\to 1$ corresponding to the Gaussian (normal) distribution as required by CLT (the same happens to the diffusion coefficient $D$)~\cite{arias2021methods}. The diffusion process in a porous medium is another example, where if the medium structure or external field changes over time, the CO fractional diffusion is not applicable~\cite{atangana2013use, hantush1955non}.  This characteristic poses fundamental challenges and introduces the need to reconsider the governing equation, considering the time dependence of these exponents. To address this limitation, using variable order (VO) fractional diffusion equations has been proposed~\cite{frame1970lassa, atangana2013use, hantush1955non}. As it has been shown in several studies that the inclusion of VO fractional derivatives can provide a more accurate representation of the underlying dynamics in many systems~\cite{li2004chaos, yang2016fractional, almeida2017caputo, atangana2015stability, alkahtani2016novel}. 
 
 This paper considers the VO fractional porous media equation (FPME) with local fractional derivative operators, where all the exponents can change slowly with time. The formalism is kept as general as possible to include a general time dependence of the exponents. By proposing a separable form for the solutions, we identify an important class of solutions that yield the ordinary $q$-Gaussian solution in the static (CO) limit. These solutions are not self-similar (SS), but the self-similarity is retrieved once we take the CO limit. In the second part of the paper, we relate these solutions to the PDF of price return in the traditional stock markets and the cryptocurrency. We assess the VO $q$-Gaussian function and inspect how this system approaches the normal diffusion counterparts over extended periods. \\
 
The paper is organized as follows: the constant order fractional diffusion process (CO) will be presented in the following section. 
Section~\ref{SEC:1} is devoted to the time-dependent (VO) exponents and their solutions with and without drift. The application to the stock markets is studied in Section~\ref{SEC:Application}.

\section{ Constant Order (CO) fractional diffusion process}
The anomalous diffusion is a diffusion process with a non-linear relationship between the mean squared displacement and time with an anomalous diffusion exponent $\alpha$. For any $d$-dimensional space, it is characterized by the following scaling relation
\begin{equation}
R(t)\equiv\sqrt{\left\langle r^2(t)\right\rangle}\propto t^{H} ,
\label{Eq:scaling0}
\end{equation}
where $r(t)$ is the end to end distance at time $t$, $\left\langle ...\right\rangle$ is the ensemble average and $H=1/\alpha$ is the corresponding Hurst exponent. For a normal diffusion $\alpha=2$, while for the super- (sub-) diffusion $\alpha<2$ ($\alpha>2$). The anomalous diffusion can be due to time correlations, as well as the fractal structure of the \textit{space}. An important primitive example of anomalous diffusion was given by Havlin and Ben-Avraham for random walks on fractal objects, the PDF of which is given by~\cite{havlin1987diffusion}
\begin{equation}
    P(x,t)\propto R(t)^{-d_f}\exp \left[-c\left(\frac{x}{R(t)}\right)^{\frac{\alpha}{\alpha-1}}\right],
    \label{Eq:example}
\end{equation}
where $d_f$ is the fractal dimension of the space in which the random walker is doing an exploration process. Other types of distributions with the same scaling relation between $x$ and $R(t)$ are proposed to describe anomalous diffusion processes in different physical systems, which are special solutions of the modified Fokker-Planck equations (FPEs). These modifications of the FPE may include the fractionalization of the space as well as the time derivative operators, and non-linearization depending on the system that FPE is going to describe. Various fractional diffusion equations have been introduced each of which has its own advantages and weaknesses~\cite{umarov2009variable, magin2008anomalous,chechkin2005fractional}. A fractionalization of derivative can be either local or non-local depending on the (temporal and spatial) nature of the system~\cite{arias2022testing}. The examples are the Schneider and Wyss time-derivative fractionalization~\cite{schneider1989fractional}, O'Shaugnessy and Procaccia space-derivative fractionalization~\cite{o1985analytical}, Giona and Roman space-time-derivative fractionalization~\cite{giona1992fractional}, and  more general cases~\cite{metzler1994fractional}. \\

An important feature in Eq.~\ref{Eq:example} is related to its scaling behavior. This equation suggests that for a $d$-dimensional system, $R=|\textbf{x}|$ (where $\textbf{x}$ shows the position of a random walker) scales with a general function of time $\phi(t)$, so that 
\begin{equation}
\textbf{x}\to \lambda \textbf{x}\ , \ \phi(t)\to \lambda\phi(t)\ , \ P\to\lambda^{-d}P\ ,
\end{equation}
(see Appendix~\ref{App:dimensional} for more details). Here $\phi(t)$ is a time-dependent function characterizing the anomalous diffusion. This suggests the following scaling solution
\begin{equation}
P(\textbf{x},t)=\frac{1}{ \phi(t)^d}  F  \left[\frac{\textbf{x}}{\phi(t)} \right].
\label{Eq:generalForm}
\end{equation} 
The  nature of anomalous diffusion is directly calculated using
\begin{equation}
    R^2=\left\langle r(t)^2\right\rangle\propto\int \text{d}^d\textbf{x}|\textbf{x}|^2F\left[\frac{\textbf{x}}{\phi(t)} \right]\propto \phi(t)^2. 
    \label{Eq:scaling}
\end{equation}
The scaling properties of the time series are associated with the form of $\phi(t)$. In fact, for the solution of Eq.~\ref{Eq:scaling0} we have $\phi(t)=\phi_{\text{SS}}(t)$ where the index ``SS'' points out the self-similarity law, given by~\cite{gharari2021space,alonso2019q}:
\begin{equation}
\begin{split}
\phi_{\text{SS}}(t)\propto t^{1/\alpha}.
\end{split}
\label{eq:scale}
\end{equation}
Combining Eq.~\ref{Eq:scaling} and Eq.~\ref{eq:scale}, one reaches  Eq.~\ref{Eq:scaling0}.\\

An important well-known example is the fractional Brownian motion, in which the PDF follows a Gaussian distribution with a self-similar structure. For a good review see Appendix~\ref{App:FBM}
and~\cite{beran1994statistics,mandelbrot1968fractional}.
The Levy-stable distribution is another example of a self-similar system that has vast applications in stochastic processes, including the stock markets. The heavy-tail behavior observed in stock market price fluctuations has been a cornerstone for many scientists in supporting the use of Levy-stable distributions for modeling the stochastic behavior of the price return~\cite{bielinskyi2019detecting, chang2020dynamic, montillet2015modeling, bielinskyi2019levy, scalas2006art, bucsa2014practical, arias2021methods}. \\
There are, however, some pieces of evidence indicating that Levy-stable distributions are not sufficient to describe the stylized facts of the price return. Recent observations of PDF of price returns for the S\&P500 have shown that they follow more general distributions ~\cite{alonso2019q}.
The Levy-stable distribution provides only an estimation of the stock market fluctuations at low frequencies where the correlations can be neglected. However, correlations during the first
minutes on the price fluctuations were observed at high frequencies, making the Levy regime no longer applicable.
Additionally, the characteristic exponents used to model the power-law tails in the PDF of price returns during the first minutes lie outside the Levy regime \cite{gharari2021space}. This divergence highlights a common occurrence in the modeling of complex systems, which can be attributed to the nonlinear nature of the governing physical phenomena. 

In nonlinear systems, the principles of homogeneity and superposition do not hold. These systems exhibit a distinctive property known as \textit{non-extensivity}, meaning that their corresponding entropy is not additive. A notable example of non-extensive systems is observed in the porous media equation (PME), which possesses broad applications in stochastic processes, including the analysis of stock markets. More accurate models can be created by introducing a fractional version of the PME. This extension offers a powerful theoretical framework with the potential to effectively describe a wide range of stochastic systems.
The local (Katugampola) fractional PME reads~\cite{gharari2021space}
\begin{equation}
\begin{split}
&\dfrac{\partial^{\xi}}{\partial t^{\xi}}P(x,t)=D\dfrac{\partial^{2}}{\partial x^{2}}P(x,t)^{\nu},
\end{split}
\label{eq:f}
\end{equation}
where $\xi$, $ D$, and $\nu\equiv 2-q$ are the constant parameters to be found by fitting the data with the time series under investigation. $D$ is the diffusion coefficient in the limit $q, \, \xi\rightarrow 1$ where the normal diffusion is retrieved.
Eq~\ref{eq:f} admits solutions in terms of $q$-Gaussian and generalized $q$-Gaussian functions~\cite{alonso2019q, gharari2020local}, which forms an important class of functions with a wide range of applications~\cite{alonso2019q,borland2002option}, shown as
\begin{equation}
P(x,t)=\frac{1}{C_q \phi_{\text{SS}}(t)}  e_q  \left[ \left(\frac{x}{\phi_{\text{SS}}(t)} \right)^2\right],
\label{Eq:generalForm1}
\end{equation} 
where $e_q(x)\equiv \left(1-(1-q)x^2\right)^{\frac{1}{1-q}}$ is a generalization of the exponential function, and $C_q$ is a $q$-dependent normalization factor. The self-similar time part reads:
\begin{equation}
\begin{split}
\phi_{\text{SS}}(t)\equiv(D' t)^{1/\alpha}.
\end{split}
\label{eq:m2}
\end{equation}
where the parameter $\alpha=\frac{3-q}{\xi}$ is the anomalous diffusion exponent associated with the self-similarity of the time series and $ D'\equiv (D/\xi)^{1/\xi}$ is the modified diffusion parameter to be estimated using the real data analysis. The evolution equation of the price return's PDF can be constructed based on the $q$-Gaussian fitting. Originally conceived for studying fluid propagation in porous media, PME has significantly broadened its scope over time. Now PME is used to investigate any diffusion process where the diffusion coefficient depends on the state variable, the most important of which is the stock market with $q$-Gaussian PDF~\cite{alonso2019q,adams1992field}. In our previous studies, we investigated the fractional PME with local and non-local fractional derivatives, focusing solely on its solutions for describing the PDF of S\&P500 market index. Considering both cases, the results obtained from the non-local derivatives were found to be more accurate~\cite{gharari2020local}.\\
Despite the fact that a generalized form of $q$-Gaussian PDFs better describes the PDFs of S\&P500 data at any time, the exponents have been shown to vary over time~\cite{arias2021methods}. 
Solving Eq.~\ref{eq:f} with ``time-dependent exponents" introduces a contradiction because the time dependence of the exponents should have been considered in the governing equation from the outset. Such governing equations with time-dependent exponents are referred to as ``variable order" (VO) governing equations. The very important question we should answer is: \textit{Which is the generalized non-linear fractional Fokker-Planck equation that governs the probability density function (PDF) with variable orders? }

Variable exponents are observed in complex diffusion processes ~\cite{fedotov2012subdiffusive,wang2019wellposedness,liu2016maximum,fang2020fast,kian2018time}.
The diffusion properties of homogeneous media are usually modelled by constant order (CO) time-fractional diffusion processes, for example, see \cite{carcione2013theory}. However in complex media where heterogeneous regions are present the CO fractional dynamic models are not robust over long time scales. Additionally, when considering diffusion processes in porous media where the medium structure or external field changes with time, the use of CO fractional dynamic models may not yield satisfactory results~\cite{atangana2013use, hantush1955non}. In such cases, the variable order (VO) time-fractional model emerges as a more suitable approach for describing space-dependent anomalous diffusion processes~\cite{sun2009variable}. Previous works on VO diffusion models have made substantial contributions to the modeling and analysis of complex systems
\cite{fedotov2012subdiffusive, frame1970lassa,wang2019wellposedness, kian2018time,fang2020fast, liu2016maximum}. Building upon these works, this paper aims to generalize the PME by incorporating variable exponents. The investigation focuses on systematically exploring the problem with variable exponents and solving a time variable order porous media equation (VO-PME).
The VO-PME reads
\begin{equation}
\begin{split}
&\dfrac{\partial^{\xi(t)}}{\partial t^{\xi(t)}}P(x,t)=D(t)\dfrac{\partial^{2}}{\partial x^{2}}P^{\nu(t)}(x,t),
\end{split}
\label{eq:d_e_x1}
\end{equation}
where $ \xi (t)$,  $D(t)$, and $\nu(t)\equiv 2-q(t)$ now vary with time. Some properties of the CO equations cannot be extrapolated to the VO counterpart. For example, While one may be tempted to derive the effective time-dependent Hurst exponent as $H(t)\equiv \frac{\xi(t)}{3-q(t)}$, it is crucial to exercise caution when utilizing this expression. The reason is that the definition of the Hurst exponent is based on the autocorrelation function, which has not been explicitly obtained for the VO-PME in this study. Therefore, the aforementioned expression should be interpreted with care. In the following sections, we find an important class of solutions for the VO-PME. 

\section{A  local variable order non-linear time  diffusion equation}\label{SEC:1} 
In this section, we consider a time-dependent VO-PME as follows:
\begin{equation}
\begin{split}
&\dfrac{\partial^{\xi(t)}}{\partial t^{\xi(t)}}P(x,t)=D(t)\dfrac{\partial^{2}}{\partial x^{2}}P^{\nu(t)}(x,t),
\end{split}
\label{eq:d_e_x}
\end{equation}
where $\xi(t)$ and $\nu(t)=2-q(t)$ are VO exponents and $D(t)$ is a slow-varying time-dependent diffusion coefficient. In this equation the time derivative is fractionalized using a Katugampola derivative, see Appendix~\ref{App:Katugampola-derivative} for the details. Note that in the limit $\xi,\nu, D$ are constant, the solution given by Eq.~\ref{Eq:generalForm1} is retrieved, i.e. the $q$-Gaussian distribution. In the analogy of Eq.~\ref{Eq:generalForm} ($d=1$), we consider the factorized solution $P(x,t)=\frac{1}{\phi(t)} F(\frac{x}{\phi(t)} )$. This approach enables us to use the method of separating variables, where $\phi(t)$ satisfies a time-fractional equation. By inserting Eq.~(\ref{Eq:generalForm}) into Eq.~(\ref{eq:d_e_x}), we find that (also see Appendix~\ref{App:Katugampola-derivative})
\begin{equation}\label{equ.13}
-\frac{\phi^{\nu(s)}(s)}{D(s)}\dfrac{\partial^{\xi(s)}\phi(s)}{\partial s^{\xi(s)}}=\left( \frac{d}{dz}[ zF]\right)^{-1} \frac{d^{2}}{dz^{2}}F^{\nu(s)},
\end{equation}
where we change the variable $(x,t)\to (z\equiv \frac{x}{\phi(t)},s\equiv t)$. To simplify the calculations, we assume that the function $q(s)$ is a slow-varying function so that the derivatives of $q(s)$ with respect to $s$ can be neglected as a first-order approximation. Thus, the right-hand side is a sole function of $z$, while the left-hand side is a sole function of $s$. Then we find
\begin{equation}
\left\lbrace \begin{matrix}
\frac{\phi^{\nu(s)}(s)}{D(s)}\dfrac{\partial^{\xi(s)}\phi(s)}{\partial s^{\xi(s)}}=k  &  $(I)$\\
\\
\left( \frac{d}{dz}[ zF]\right)^{-1}  \frac{d^{2}}{dz^{2}}F^{\nu(s)}=-k &  $(II)$
\end{matrix}\right. 
\label{Eq:separation}
\end{equation}
where $k$ is a real number, which serves as a free parameter. Using the properties of VO-K derivative (K stands for Katugampola, see
 Appendix~\ref{App:Katugampola-derivative})
we find that
\begin{equation}
\dfrac{\partial^{\xi(s)}\phi(s)}{\partial s^{\xi(s)}}=\frac{1}{s^{\xi(s)-1}}\phi'(s)
\end{equation}
where (here and throughout of the paper) $f'(s)$ shows the first derivative of $f(s)$ with respect to the argument $s$. Therefore, Eq.~\ref{Eq:separation}-(I) leads to:
\begin{equation} 
  \phi^{\nu(s)}(s)\phi^{\prime }(s) =k D_ss^{\xi(s)-1}.
  \label{Eq:phi-VO}
\end{equation} 
To solve this equation, taking a similar approach from Eq.~\ref{eq:scale}, we assume:
\begin{equation}
\phi(s)=\phi_0 s^{1/\tilde{\alpha}(s)},
\label{Eq:alpha}
\end{equation}
where $\tilde{\alpha}(s)$ is a nea slow-VO exponent, and $\phi_0$ is a constant. Note that in the constant order case $\tilde{\alpha}(s)$ is identical to $\alpha$. Substituting this into Eq.~\ref{Eq:phi-VO} we find ($\nu_s\equiv \nu(s)$, $\xi_s\equiv \xi(s)$, $\tilde{\alpha}_s\equiv \tilde{\alpha}(s)$ and $D_s\equiv D(s)$)
\begin{equation}
\phi_0^{\nu_s+1}s^{\frac{\nu_s+1}{\tilde{\alpha}_s}}\frac{d}{ds}\left(\frac{\ln s}{\tilde{\alpha}_s}\right)=kD_ss^{\xi_s-1},
\end{equation}
or in terms of a new variable $y\equiv (\nu_s+1)\frac{\ln s}{\tilde{\alpha}(s)}$ (so that $e^y=\left(\frac{\phi(s)}{\phi_0}\right)^{\nu_s+1}$) we have:
\begin{equation}
(\nu_s+1)e^y\frac{d}{ds}\left(\frac{y}{\nu_s+1}\right)=k\tilde{D}_ss^{\xi_s-1},
\end{equation}
where,
\begin{equation}
    \tilde{D}_s\equiv \frac{(\nu_s+1)}{\phi_0^{\nu_s+1}}D_s.
\end{equation}
When $\nu_s$ is a smooth function of $s$, by considering that $\nu_s=2-q_s$, we can ignore its first derivative, so one can easily cast the equation to the form:
\begin{equation}
\frac{d}{ds}e^y=k\tilde{D}_ss^{\xi_s-1},
\end{equation}
\and the solution is  with initial condition $y_0\equiv y(s_0)$ is:
\begin{equation}
e^y=e^{y_0}+kG(s,s_0).
\label{Eq:phisol}
\end{equation}
In Eq. \ref{Eq:phisol} we define:
\begin{equation}
\begin{split}
G(s,s_0) &\equiv \int_{s_0}^s \tilde{D}_tt^{\xi_t-1}dt\equiv \int_{s_0}^s g(t)dt,
\end{split}
\label{Eq.g_t}
\end{equation}
where,
\begin{equation}
g(t)\equiv \frac{D_t(\nu_t+1)}{\phi_0^{\nu_t+1}}t^{\xi_t-1}.
\label{Eq.Dt_para}
\end{equation}
Equation~\ref{Eq:phisol} can be written in the following form:
\begin{equation}
\phi(s)=\phi_0\left[k_1+kG(s,s_0)\right]^{\frac{1}{\nu_s+1}}.
\label{Eq:phiFinal}
\end{equation}
where $k_1\equiv \left(\frac{\phi(s_0)}{\phi_0}\right)^{\nu_{s_0}+1}$. Note that, for the solution to be real, we should always have the condition:
\begin{equation}
    G(s,s_0)\ge -\frac{k_1}{k}.
    \label{Eq:positivityg}
\end{equation}
By choosing 
\begin{equation}
k>0,
\label{Eq:kPositive}
\end{equation}
we see that this condition is satisfied given that $k_1\ge 0$.
Note $k_1$ is a location parameter, and it is related to the initial condition, i.~e. it sets the initial width of the PDF. When $k_1=0$, the PDF is initially the Dirac delta function, and the solution of the equation will correspond to the Green Function. In applications where the initial condition is not given, we can set $k_1$ to zero.  The constant $k$ is a scale parameter so that without loss of generality we can set $k=1$.
Equations~\ref{Eq:alpha} and~\ref{Eq:phiFinal} the exponent $\tilde{\alpha}(s)$ is obtained as:
\begin{equation}
\frac{1}{\tilde{\alpha}_s}=\frac{\ln\left[k_1+kG(s,s_0)\right]}{(\nu_s+1)\ln s}.
\label{Eq.g_s}
\end{equation}
The initial time $s_{0}$ can be set to zero. One can easily demonstrate that, when considering fractional constant exponents, $\nu_s=\nu,$ $\xi_s=\xi $, Eq.~\ref{Eq:phiFinal} coincides with the result of ordinary PME, that is 
\begin{equation}
\tilde{\alpha}_s\to\alpha,\  \phi (s)\to \phi_{\text{ss}} (s)  
\end{equation}
which is given in Eq.~\ref{eq:m2}. Specifically, in the limit $D_s\rightarrow  D=\text{const.}, \nu_s\rightarrow 1$,  $\xi_s\rightarrow 1$ and $k_1=0$, one retrieves the  normal diffusion (ND), for which 
\begin{equation}
   \phi(s)\to \phi_{\text{SS}}^{(ND)}\equiv a s^{\alpha_{\text{ND}}}, \ \tilde{\alpha}_s\to\alpha_{\text{ND}}=\frac{1}{2}  
\end{equation} 
where $a=k\sqrt{2D}$.\\

In the next step, we find the solution of $F$. From now on we set $k=1$ and $k_1=0$, bearing in mind that the formulas can be generalized by considering other values of these parameters. We recall  Eq.~(\ref{Eq:separation})-II,
\begin{equation}\label{eq.s2}
\frac{d}{dz}[ zF]=- \frac{d^{2}}{dz^{2}}F^{\nu_s}.
\end{equation}
Let us consider the following trial special solution as a standard form:
\begin{equation}\label{Eq:trial0}
 F(z,s)=  (c+\eta_sz^{2})^{^{ \frac{1}{\nu_s-1}}}
\end{equation}
 where $c$ is a constant, and $\eta_s$ is a pure function of $s$ to be determined. Before going into the details, let us comment on the real-positively of this solution. To guarantee this, one has to impose:
 \begin{equation}
c+\eta_sz^2\ge 0.
\label{Eq:etapositivity}
 \end{equation}
 This inequality is satisfied only when
  \begin{equation}
\eta_s\ge 0,
\label{Eq:inequality}
 \end{equation}
 and at the same time $c\ge 0$, which we arbitrarily set it to $c=1$ (which can be done using a normalization). For the trial solution Eq.~\ref{Eq:trial0}, one has
\begin{equation}
\frac{d }{dz }[F^{\nu_s}]=\frac{2 \nu_s \eta_s}{\nu_s-1}z F.
\end{equation}
After incorporating this expression into Eq.~(\ref{eq.s2}), we obtain
\begin{equation}
\eta_s=\dfrac{1- \nu_s }{2\nu_s},
\end{equation}
which completes the solution. After all, the Eq.~\ref{Eq:inequality} has to be satisfied, for which we should satisfy the following inequality
\begin{equation}
\begin{split}
&\frac{q_s-1}{2-q_s}\ge 0.
\end{split}
\end{equation}
Noting that
\begin{equation}
 \frac{q_s-1}{2-q_s} \left\lbrace \begin{matrix}
 \ge 0 & \text{if} \ 1\le q_s< 2
 \\
< 0 & \text{if} \ q_s< 1 \ \text{or} \ q_s> 2,
\end{matrix} \right.    
\label{ineq:branches}
\end{equation}
we find a physically relevant interval where the Eq.~\ref{Eq:etapositivity} is fulfilled: $1\le q<2$. Outside of this range, the solutions become imaginary. Finding another set of solutions is beyond the scope of the present paper as the study cases of price return in the stock market are restricted to the interval $1<q\le 2$; therefore, the lower branch in Eq~\ref{ineq:branches} is applicable. Note that for $k\ne 1$ we should satisfy $\frac{q_s-1}{k(2-q_s)}\ge 0$, which is automatically satisfied for $1\le q<2$ given the Eq.~\ref{Eq:kPositive}.\\

Altogether, we realize that $F(z)$ and $\phi (t)$ are positive functions, and eventually:
\begin{equation}
P(x,t)\propto\frac{A_{q}(t)}{  \phi (t)}  \left[1 +\eta_t  \left(\frac{x}{\phi (t)}\right)^2 \right]^{\frac{1}{\nu_t-1}},
\label{Eq:P_x_t_l}
\end{equation}
where $A_{q}(t)$ is a normalization factor (independent of $x$). Now defining $ \eta_{q_t} =\frac{\eta_t}{q_t-1}=\dfrac{1 }{2(2-q_t)}$, we find that
\begin{align}\nonumber 
P(&x,t\vert x_{0},t_{0})= \frac{A _{q}(t)}{\phi(t)} e_{q_t}\left[-\eta_{q_t}\left(\frac{x}{\phi(t)}\right)^2\right],
\label{Eq:12}
\end{align}
where
\begin{equation}
e_q[x]\equiv [1+(1-q)x]^{\frac{1}{1-q}},
\end{equation}
is a $q$-Gaussian function. To make the notation more abstract, we define:
 \begin{equation} 
  \Phi(t)\equiv \frac{\phi(t)}{\sqrt{\eta_{q_t}}}. 
 \end{equation}
Using Eq.~\ref{Eq:phiFinal} we find the explicit form
  \begin{equation} \label{eq:Phi}
  \Phi(t)=\sqrt{2(2-q_t)}\left(\int_{t_0}^tg(s)ds\right)^{\frac{1}{3-q_t}}.  
 \end{equation}
Then, $P(x,t)$ is defined as:
 \begin{equation}
P(x,t)= \frac{1}{C _{q}(t)\Phi(t)} e_{q_t}\left[-\left(\frac{x}{\Phi(t)}\right)^2\right],
\label{Eq:PDF}
\end{equation}
where,
 \begin{equation} 
 C_{q}(t)=\sqrt{\eta_{q_t}}A_q^{-1}(t)=\frac{\sqrt{\pi}\Gamma\left(\frac{3-q_t}{2(q_t-1)}\right)}{\sqrt{(q_t-1)}\Gamma\left(\frac{1}{q_t-1}\right)}
 \label{eq:Cq}
 \end{equation}
and $ \Gamma(.) $ is the Gamma function. Note that the standard deviation is:
 \begin{equation}
\sqrt{\left\langle x^2\right\rangle}=\Phi(t).
\label{eq:second_moment}
 \end{equation}
 If we represent 
 \begin{equation}
\Phi(t)\equiv \Phi_0 t^{\frac{1}{\alpha_t}}
\label{Eq:alpha_real},
 \end{equation} 
 then using Eq.~\ref{Eq:alpha} one finds
 \begin{equation}
     \frac{t^{1/\tilde{\alpha}_t}}{\sqrt{\eta_{q_t}}}=Ct^{1/\alpha_t}\rightarrow \frac{1}{\alpha_t}\ln t=\frac{1}{\tilde{\alpha_t}}\ln t-\frac{1}{2}\ln \left(\eta_{q_t}\right)-\ln C.
     \label{Eq.alpha_tilde}
 \end{equation}
 where $C\equiv \frac{\Phi_0}{\phi_0}$. Using Eq.~\ref{Eq.alpha_tilde}, we can calculate $\alpha_t$ if $\tilde{\alpha}_t$ and $\eta_t$ are provided and vice versa. In the practical situations, one calculates $\alpha_t$ using Eq.~\ref{Eq:alpha_real}, and the $\tilde{\alpha}_t$ can be obtained from Eq.~\ref{Eq.alpha_tilde}. A similar formulation for the case with drift is presented in  Appendix~\ref{SEC:drift}. \\
 
\begin{figure*}
{\includegraphics[scale=0.55,trim=2.0cm  0.4cm 0cm 0cm, angle =0 ]{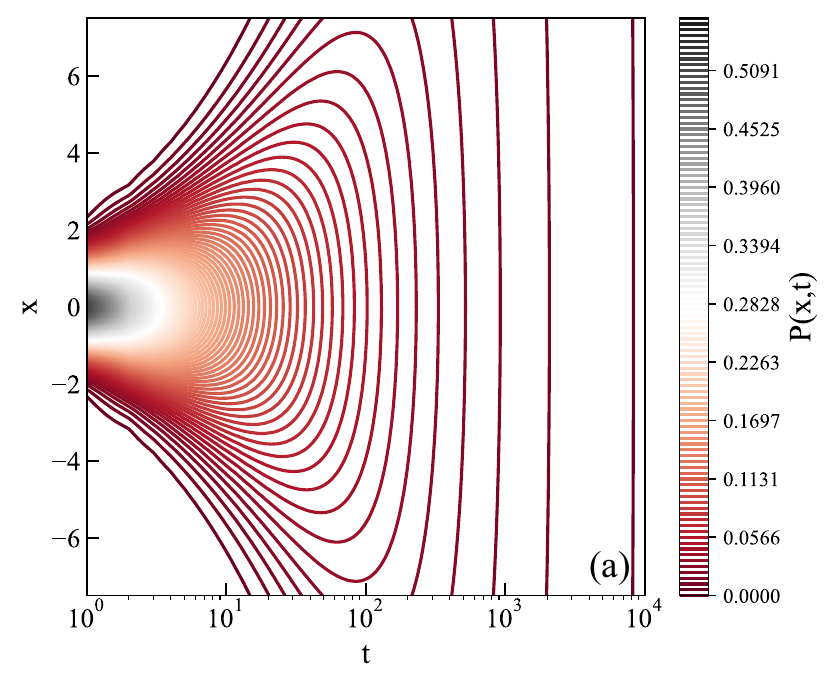}}
{\includegraphics[scale=0.46,trim=0cm  0cm 0cm 0cm, angle =0 ]{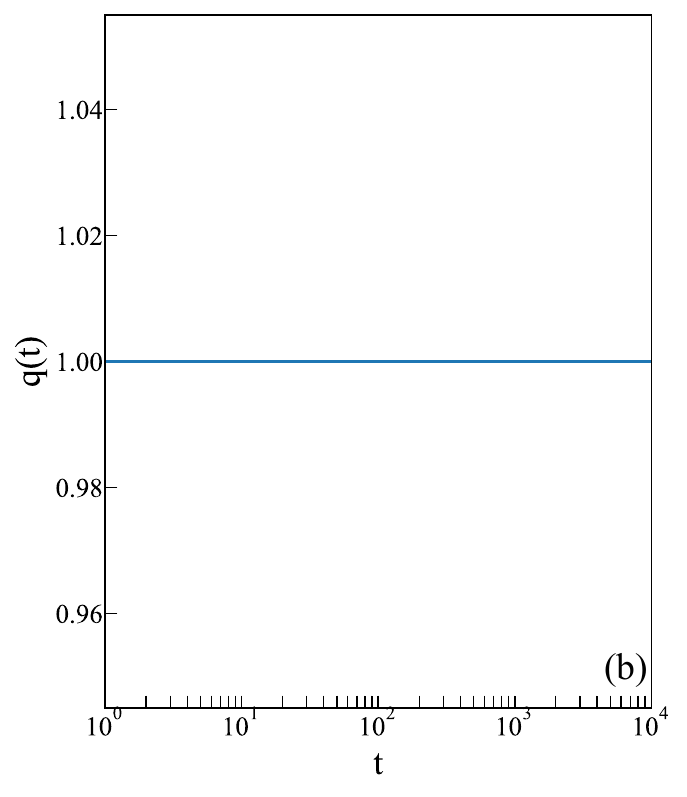}}
{\includegraphics[scale=0.46,trim=0cm  0cm 0cm 0cm, angle =0 ]{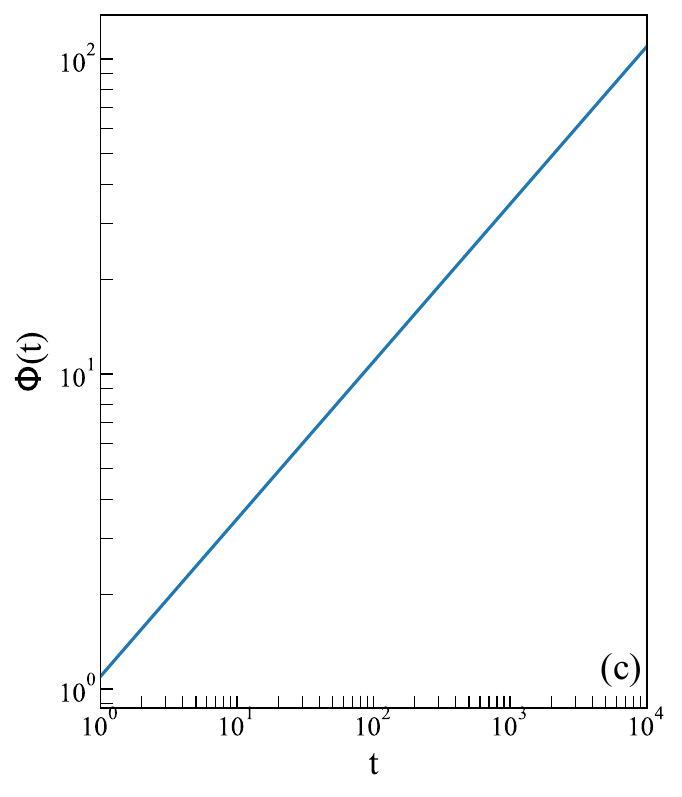}}

{\includegraphics[scale=0.55,trim=2.0cm  0.4cm 0cm 0cm, angle =0 ]{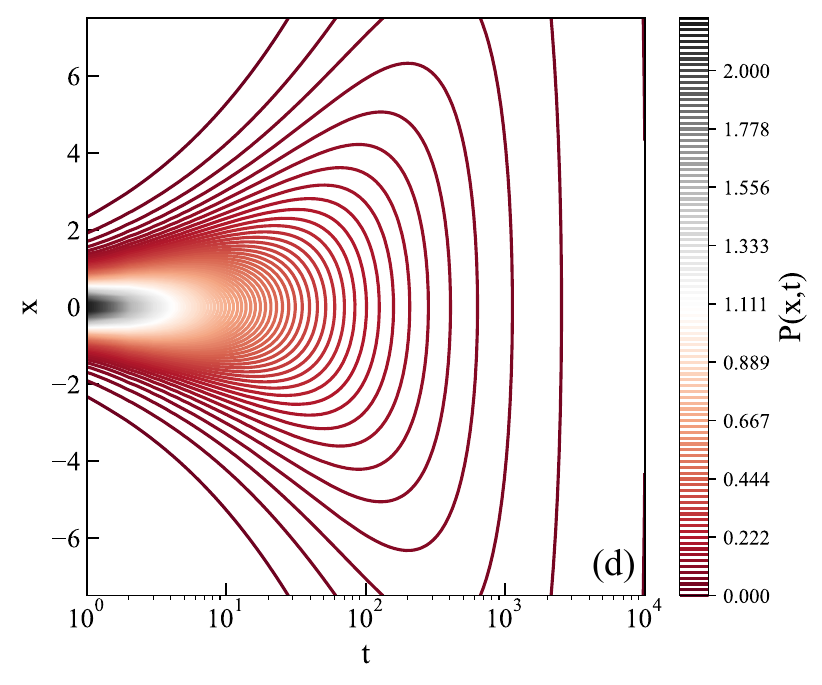}}
{\includegraphics[scale=0.46,trim=0cm  0cm 0cm 0cm, angle =0 ]{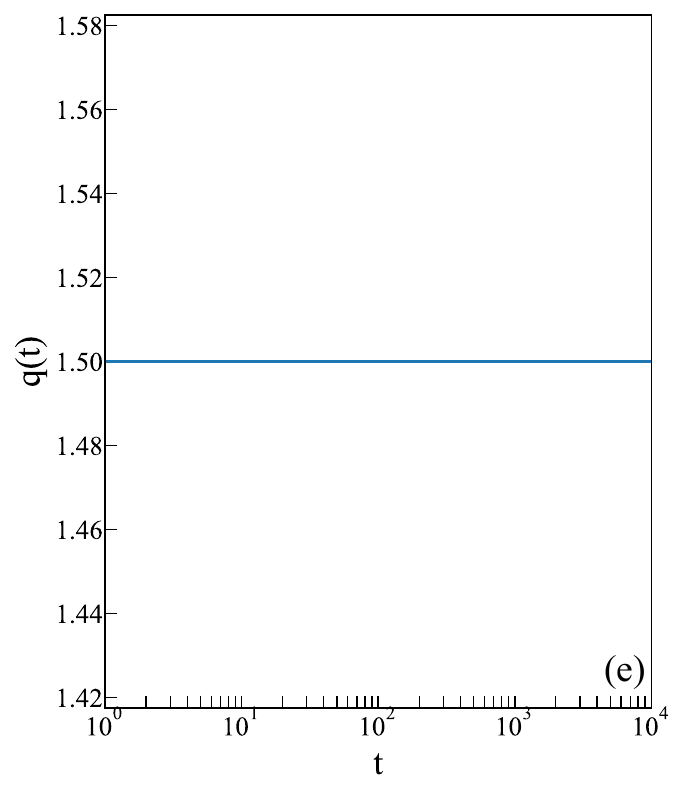}}
{\includegraphics[scale=0.46,trim=0cm  0cm 0cm 0cm, angle =0 ]{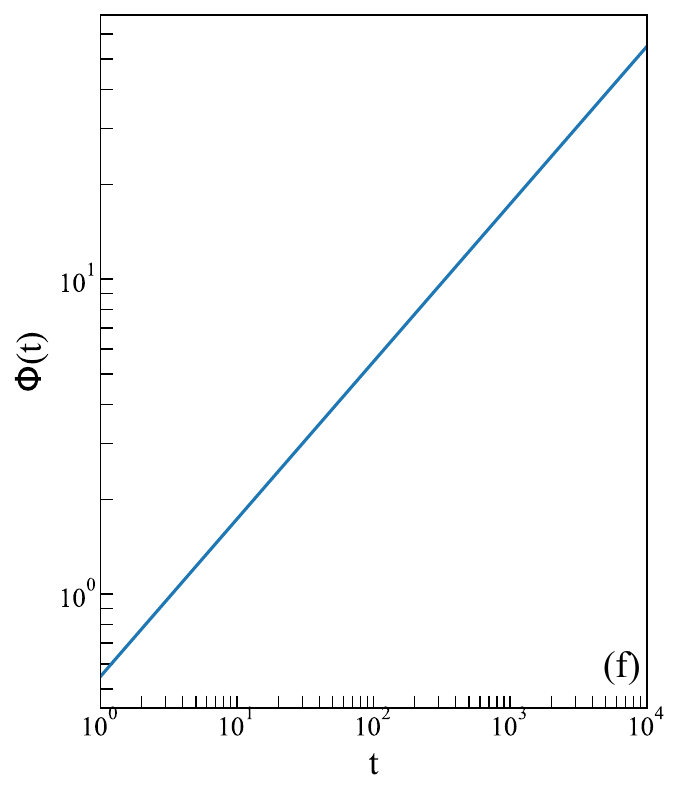}}

 {\includegraphics[scale=0.55,trim=2.0cm  0.5cm 0cm 0cm, angle =0 ]{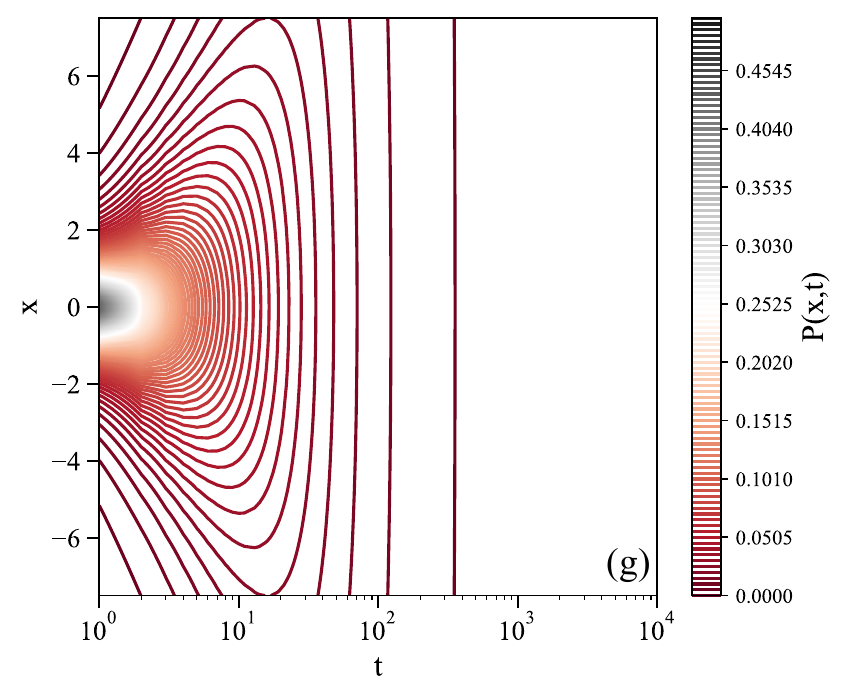}}
{\includegraphics[scale=0.46,trim=0cm  0cm 0cm 0cm, angle =0 ]{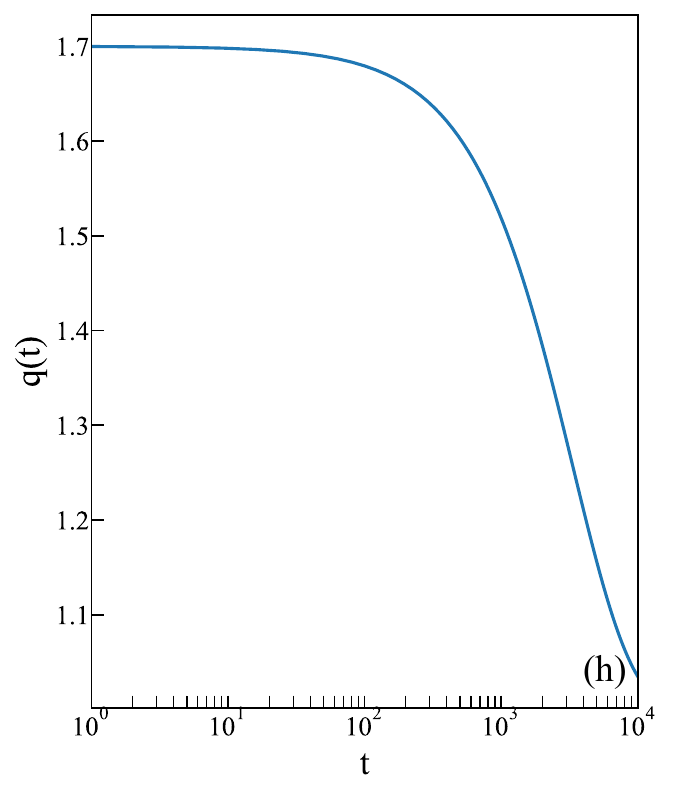}}
{\includegraphics[scale=0.46,trim=0cm  0cm 0cm 0cm, angle =0 ]{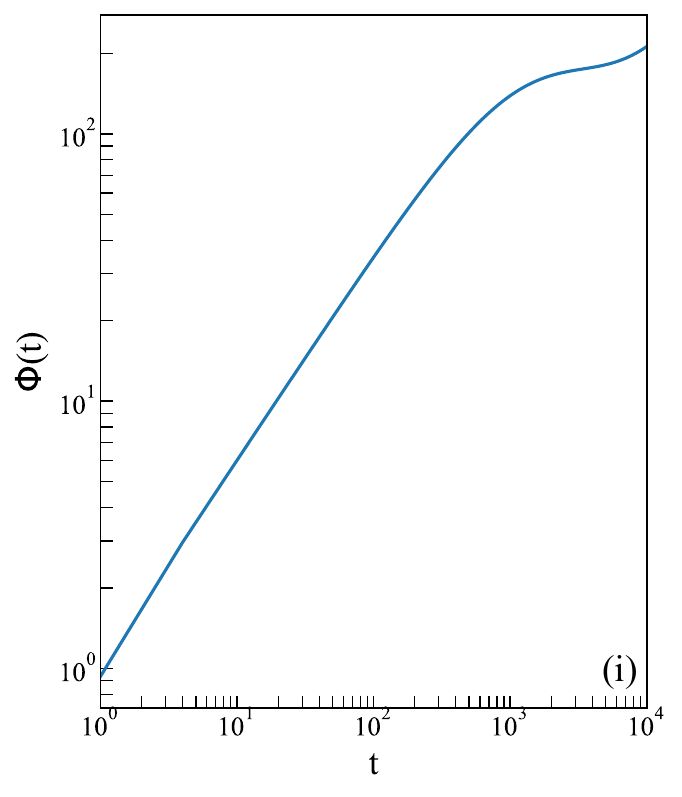}}
\caption{\label{fig:1} Summary of PMEs' solutions. (a) The time evolution of the diffusion equation or Fick's second law is described by $q=1$ and $\Phi=\sqrt{4Dt}$ presented at (b) and (c), respectively. (d) Time evolution of $P(x,t)$ with respect to time and space for the PME equation presented in  Eq.~\ref{eq:f} for $q=1.50$  and $\Phi=(Dt)^{1/\alpha}$ shown at  (e) and (f), respectively. (g) Time evolution of VO $q$-Gaussian for a $q(t)=(q_0-1)e^{-at}+1$ with $a=0.0003$ and $q_{0}=1.7$, at panel (h). Panel (i) displays the variation of $\Phi(t)$ according to Eq.(\ref{eq:Phi}).}

\end{figure*}

 In the rest of this section, we provide some results for the VO $q-$Gaussians for various functions $q(t),\,\xi(t)$ and
$D(t)$. Figure \ref{fig:1} displays the behavior of Eq. (\ref{Eq:P_x_t_l}) and compares it with the solutions of normal diffusion and porous media processes, respectively. 
Figure \ref{fig:1}-a shows the solution of the diffusion equation or Fick's second law, i.e. Eq. (\ref{eq:f}) for $q=1$ and $\xi=1$ with $D=0.3$, in Figure \ref{fig:1}-b and c we can observe $q=1$ and $\Phi = \sqrt{4Dt}$ respectively. Figure \ref{fig:1}-d exhibits the solution of $P(x,t)$ with respect to time and space for a PME, see (Eq. \ref{eq:f}), applicable for $1<q<3$. For this example we use the values of $q=1.5$, $\xi=3/4$, and $D=0.3$ in Figure \ref{fig:1}e and f we can observe that $q=1.5$, remains constant, and $\Phi = (Dt)^{1/\alpha}$ with $\alpha=\frac{3-q}{\xi} $ respectively. Finally  Figure \ref{fig:1}g  represents the VO $q$-Gaussian presented in   Eq. (\ref{Eq:P_x_t_l}) and applicable for $1<q \leq 2$  with $k=1$, $k_{1}=0 $, $q_0=1.7$, $\xi_{0}=1.3$, and  $D_0=1.3$. The value of $q(t)=(q_0-1)e^{-at}+1$, with $a=0.0003$, and $\Phi(t)$, see Eq.(\ref{eq:Phi}), are shown in Figure \ref{fig:1}-h and i, respectively.

\section{Application to Stock Markets and Cryptocurrency}
\label{SEC:Application}

In this section, we apply the VO $q$-Gaussian diffusion model to describe the evolution of PDF of stock market price return. Our empirical investigation focuses on two prominent market indices: the S\&P500 stock market index and the Bitcoin cryptocurrency. The S\&P500 dataset encompasses the period from January 2\textsuperscript{nd}, 2018, to August 9\textsuperscript{th}, 2022, with a frequency of one minute.  For the Bitcoin currency, we analyze data spanning from January 1\textsuperscript{st}, 2021, to May 9\textsuperscript{th}, 2022, with data points collected at ten-minute intervals. Prior to conducting our price return analysis, we undertake a pre-processing step to remove time frames characterized by trading amounts falling below $0.10$ USD. These instances predominantly occur approximately one hour before the stock market closes, specifically observed in the context of the S\&P500. \\

The price return is defined as~\cite{arias2021methods}
\begin{equation}\label{IV_EQ1}
X(t) = I(t_{0} + t) - I(t_{0}),
\end{equation}
where $I(t)$ is the index at time $t$, and $t_0$ is some reference time. By decomposing the price return $X(t)$ into a deterministic component $\bar{X}(t)$ and a stationary fluctuating component $x(t),$ we have
\begin{equation}
X(t) = \bar{X}(t)+x(t). 
\end{equation}
The trend $\bar{X}(t)$ was obtained by calculating the moving average of the index over a specific time window $t_{w}$in the S\&P500 and Bitcoin datasets, following the methodology described in \cite{arias2022testing}. For the S\&P500 price return, a three-month optimal time window was used, while a one-week optimal time window was employed for the Bitcoin dataset. These specific time windows were carefully chosen to ensure that the fluctuations around the trend show stationary behavior. To confirm the validity of the observed stationary behavior, we experimented with different window sizes for detrending and ultimately selected the one that allowed the PDF to show the closest convergence to a Gaussian distribution for large times. The PDFs were calculated for $x(t)$ of S\&P500 and Bitcoin at the time range $t$ $\in$  $[1,\,\,47000 ]$min. By an error-minimization process, we found that the VO q-Gaussian diffusion described in Eq.~(\ref{Eq:PDF}) provides the closest match to the observed time-dependent evolution of PDF$(x,t)$ for both S\&P500 and Bitcoin. \\

\begin{figure*}[!ht]
\centering
\includegraphics[scale=0.6,trim=0cm  0cm 0cm 0cm, angle =0 ]{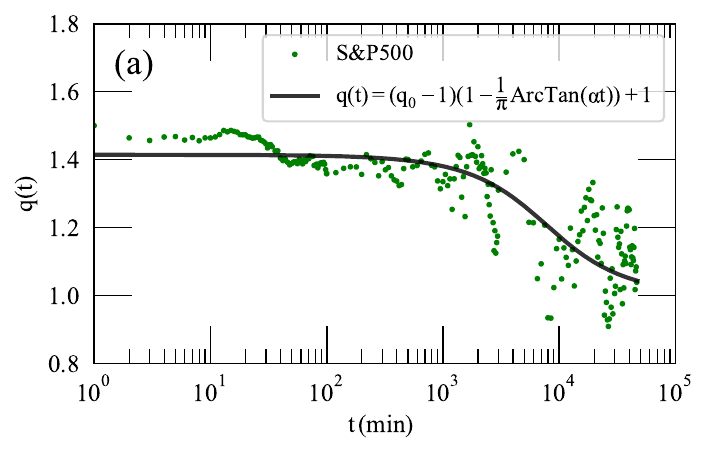}
\includegraphics[scale=0.6,trim=0cm 0cm 0cm 0cm, angle =0 ]{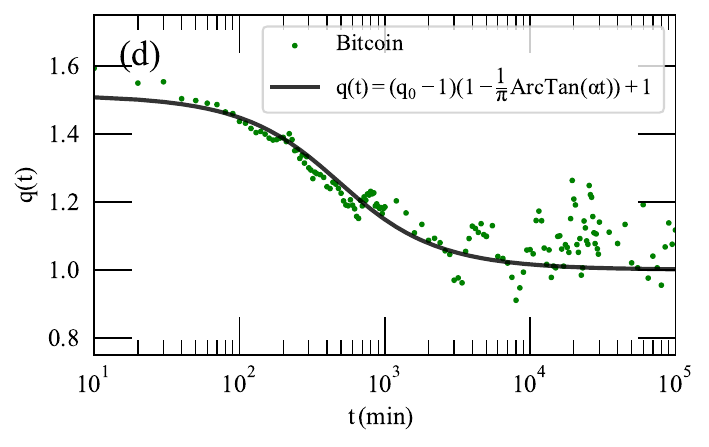}
\includegraphics[scale=0.6,trim=0cm 0cm 0cm 0cm, angle =0 ]{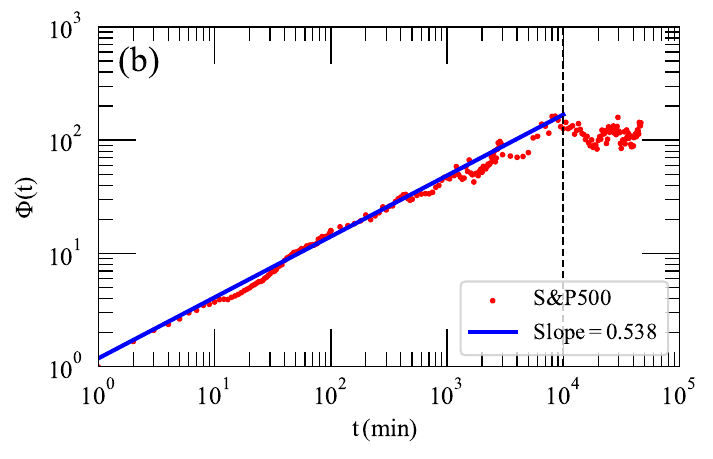}
\includegraphics[scale=0.6,trim=0cm 0cm 0cm 0cm, angle =0 ]{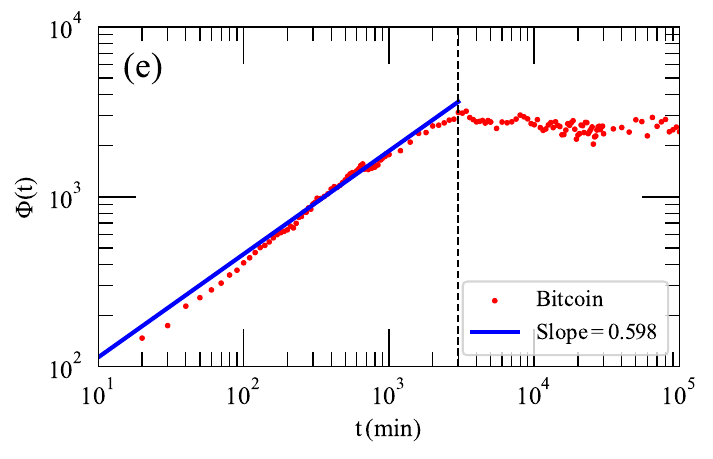}
\includegraphics[scale=0.6,trim=0cm 0cm 0cm 0cm, angle =0 ]{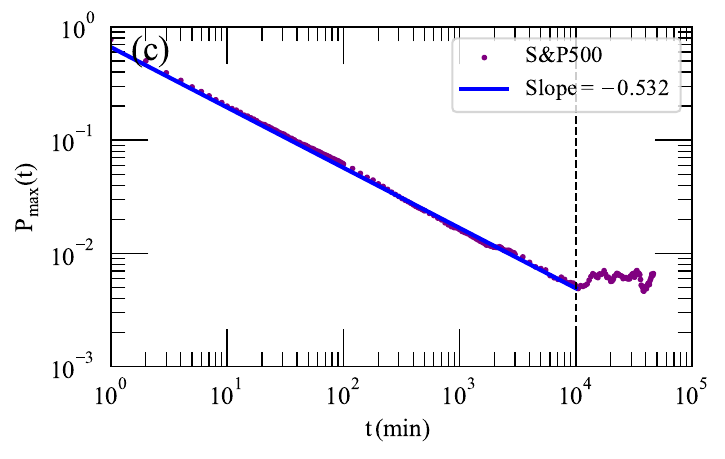}
\includegraphics[scale=0.6,trim=0cm  0cm 0cm 0cm, angle =0 ]{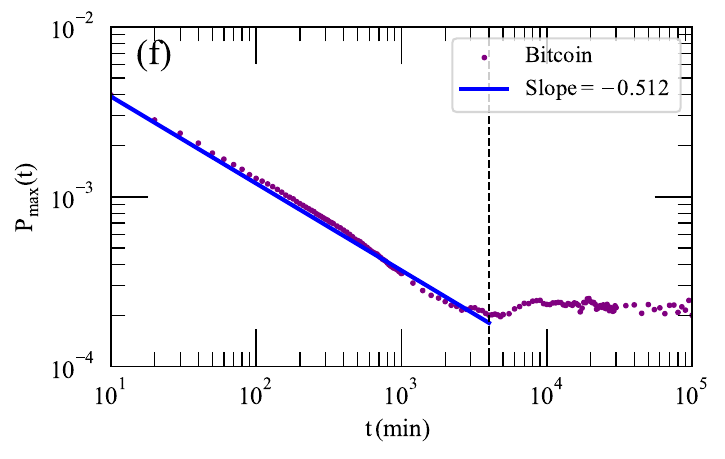}
\caption{\label{fig.VOq_fitting}Results of fitting VO q-Gaussian to the PDFs for S\&P500 and Bitcoin. (a) Fitting parameter $q(t)$ for S\&P500, which presents a constant of 1.5 initially and slowly converges to 1. (b) Fitting parameter $\Phi(t)$ for S\&P500, with a slope of $1/\alpha=0.538$. (c) The peak of PDF ($P_{max}(t)$) for S\&P500 with a slope of $-0.532$. (d) Fitting parameter $q(t)$ for Bitcoin, initially at 1.5 and converges to 1 faster than S\&P500. (e) Fitting parameter $\Phi(t)$ for Bitcoin, with a slope of $1/\alpha=0.598$. (f) The peak of PDF ($P_{max}(t)$) for Bitcoin with a slope of $-0.512$.}
\end{figure*}

Figure \ref{fig.VOq_fitting} presents the results of the calibration process, displaying the functions $q(t)$, $\Phi(t)$, and $P_{\text{max}}(t)$ obtained by curve fitting the PDFs derived from the datasets to  Eq. (\ref{Eq:PDF}). We observed that the PDFs of S\&P500 and Bitcoin converge to a Gaussian distribution function as time $t$ increases, and this convergence is positively correlated with the chosen optimal time window $t_w$. Subfigures \ref{fig.VOq_fitting}-a and \ref{fig.VOq_fitting}-d present the converge of $q$ towards 1 as $t$ approaches $t_w$, indicating a Gaussian distribution function with $\sigma^2=1/2$, and $\mu=0$. The convergence to the normal distribution is expected given the fact that for large enough times, the conditions for the central limit theorem are satisfied. We observe also that $q(t)$ for S\&P500 and Bitcoin can be effectively modeled using the following relationship
\begin{equation*}
q(t)=(q_0-1) \left(1-\frac{1}{\pi}\arctan ( a t) \right)+1.
\end{equation*}
where $q_0$ represents the initial value of $q$ at $t_0$ and $a$ is a parameter obtained through fitting. The fitting parameters for the S\&P500 are $q_0=1.414$ and $a=1.28\times10^{-4}$, while for Bitcoin, the fitted values are $q_0=1.514$ and $a=2.0\times10^{-3}$. The black curves in subfigures \ref{fig.VOq_fitting}-a and \ref{fig.VOq_fitting}-d represent the results of this fitting process. It is notable that the Bitcoin time series exhibits a closer fit to this relationship compared to the S\&P500 time series. Specifically, for S\&P500, the fitted values of $q(t)$ remain constant $q(t)=1.4$ for approximately the first three days of trading, followed by a rapid convergence to $q(t) = 1$. Conversely, for Bitcoin, the transition takes place over a shorter duration of around 16 hours, with $q(t)$ stabilizing at $q(t) = 1$ for larger values of $t$. These findings highlight the varying dynamics and characteristics between the S\&P500 and Bitcoin markets, with Bitcoin demonstrating a more pronounced adherence to the modeled relationship for $q(t)$. Subfigures \ref{fig.VOq_fitting}-b and \ref{fig.VOq_fitting}-e present the parameter $\Phi(t)$ obtained from the $q$-Gaussian fitting for S\&P500 and Bitcoin respectively. $\Phi(t)$ showcases a distinct slope and remains constant at large times. These slopes correspond to the anomalous diffusion, where $\Phi(t) \propto t^{1/\alpha(t)}$ (Eq.~\ref{eq:scale}). The average slopes obtained for S\&P500 and Bitcoin are $1/\alpha=0.54$ and $1/\alpha=0.60$ respectively, while the local slopes depend on time. 
Moreover, as time $t$ increases, a constant value for $\Phi(t)$ becomes evident. This behavior is a consequence of the detrending process applied during the analysis. 

An important feature of the VO diffusion is that the anomalous diffusion cannot be always derived from the temporal evolution of the peak of the PDF, it has been done in previous analyses of the S\&P500 index
\cite{mantegna1995scaling,alonso2019q}. 
Figure \ref{fig.VOq_fitting}-c and Figure \ref{fig.VOq_fitting}-f shows the height of the PDF of price return ($P_{\text{max}}(t)$) for S\&P500 and Bitcoin respectively. This term is obtained as the value of the PDF at $x=0$. For both stock markets, $P_{\text{max}}(t)$ exhibits a slope initially and remains constant over a large time. A linear fitting is conducted and the slopes obtained are $-0.532$ and $-0.512$ for S\&P500 and Bitcoin. This slope is consistent with the anomalous diffusion exponents for the S\&P500 but not in the Bitcoin data. In the latter case the exponent obtained from $\Phi(t)$ is $\alpha =1.86$ that does not correspond to the exponent $1.95$ obtained from $P_{\text{max}}$. This apparent discrepancy can be understood in the light of  Eq.~\ref{Eq:PDF}. There one can derive the relationship $P_{\text{max}}(x=0,t)=\frac{1}{C_q(t)\Phi(t)}$, where $C_q(t)$ is a time-dependent parameter associated with $q$ as shown in Eq.~\ref{eq:Cq}. It is noteworthy that the absolute value of the slopes in $\Phi(t)$ and $P_{\text{max}}(t)$ shows a better agreement for S\&P500 than for Bitcoin. This disparity can be attributed to the distinction in the time-dependent term $q(t)$ for both markets. In the case of S\&P500, the value of $q(t)$ remains approximately constant for a longer period than for Bitcoin. As a result, $C_q(t)$ closely approximates a constant value, leading to a relationship of $P_{max}(t) \propto 1/\Phi(t)$, which aligns with the results. Whilst for Bitcoin, where $q(t)$ varies with time, the effect of $C_q(t)$ becomes more significant.

\begin{figure*}[!ht]
\centering
\includegraphics[scale=0.6,trim=0cm  0cm 0cm 0cm, angle =0 ]{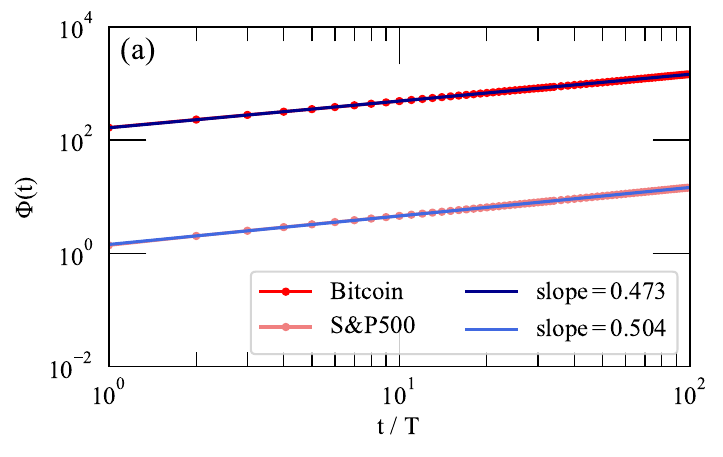}
\includegraphics[scale=0.6,trim=0cm 0cm 0cm 0cm, angle =0 ]{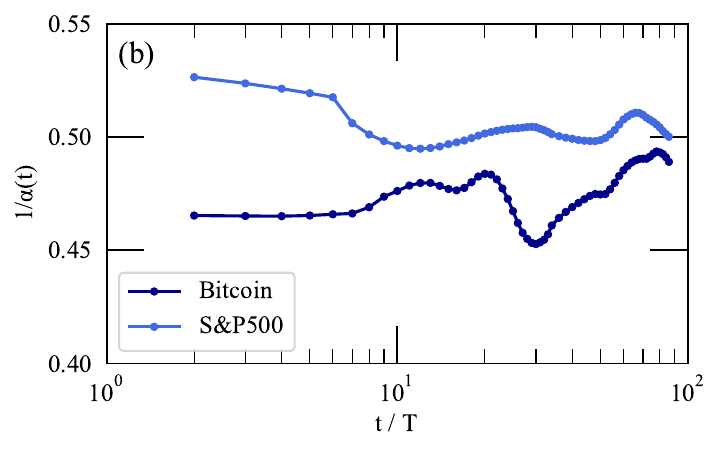}
\includegraphics[scale=0.6,trim=0cm 0cm 0cm 0cm, angle =0 ]{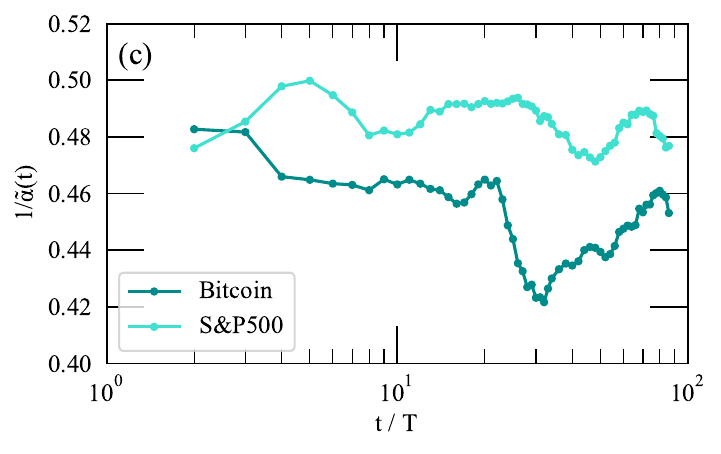}
\includegraphics[scale=0.6,trim=0cm  0cm 0cm 0cm, angle =0 ]{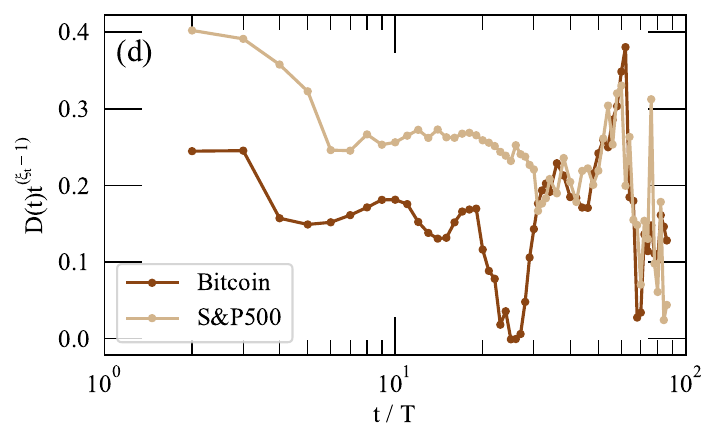}
\caption{\label{fig.VOq_alpha}(a) $\Phi(t)$ calculated for both S\&P500 and Bitcoin price return based on the second moment, both present clear slopes. (b) $1/\alpha(t)$ calculated as the localized slope of $\Phi(t)$. Both S\&P500 and Bitcoin present a convergence to 0.5, yet at a different pace. S\&P500 converges to 0.5 faster than Bitcoin. (c) $1/\tilde{\alpha}(t)$ calculated following Eq.~\ref{Eq.alpha_tilde} for S\&P500 and Bitcoin. (d) Parameter $D_{t}t^{\xi_{t}-1}$ calculated for both time series based on the relationship in Eq.~\ref{Eq.Dt_para}.} 
\end{figure*}
We now realize that neither $P_{\text{max}}(t)$ nor $\Phi(t)$ obtained from curve fitting are reliable methods to derive the exponent of the anomalous diffusion in VO diffusion processes.
In fact, the function  $\Phi(t)$ obtained in Figure~\ref{fig.VOq_fitting}-b and \ref{fig.VOq_fitting}-c present $\Phi(t)$ calculated as a fitting parameter of the PDF to the $q$-Gaussian distribution. The results of $\Phi(t)$ shown have some fluctuations and the fitting of the slope is not perfect. These fluctuations and deviations arise due to errors in the fitting process, as fitting to the q-Gaussian distribution may not always be perfect. Therefore, to get rid of systematic fluctuations it is more reliable to use the variance formula (Eq. \ref{eq:second_moment}) to estimate $\Phi(t)$ and the associated exponent $\alpha_t$ as we do in the rest of this section.

The power law relation obtained using Eq. \ref{eq:second_moment} is presented in Figure \ref{fig.VOq_alpha}-a. An adjustment was performed on the time series of both S\&P500 and Bitcoin by rescaling the price return using the data frequency ($T$). Specifically, for the S\&P500, $T= 1$ minute, while for Bitcoin, $T= 10$ minutes. 
The slopes of $\Phi(t)$ were computed for both S\&P500 and Bitcoin. The results reveal that $\Phi(t)$ for S\&P500 price return has a slope of $0.50(4)$, whereas for Bitcoin the slope of $0.47(3)$. This shows that the slope of $\Phi(t)$ is related to the type of diffusion process. These exponents however are the averaged ones over time, and the local slopes show different behaviors, i.e. they change over time which needs a ``variable order'' analysis.

We calculate the local slope of $\Phi(t)$ to determine $1/\alpha_t$ based on the aforementioned relation \ref{Eq:alpha_real}, and the results are shown in Figure \ref{fig.VOq_alpha}-b. The light-blue curve is the local slope of S\&P500, indicating superdiffusion at small time scales ($t$) with $1/\alpha(t)>0.5$, transitioning towards normal diffusion at large $t$ with $1/\alpha(t)\to 0.5$. On the other hand, the dark blue curve in Figure \ref{fig.VOq_alpha}-b represents the local slope of the Bitcoin time series, displaying subdiffusion at small times $t$ with $1/\alpha(t) < 0.5$, and gradually converging towards normal diffusion. 
Upon comparing the two curves, we observe that the S\&P500 slope exhibits a slow decrease in $1/\alpha$ at small $t$, while for Bitcoin, it remains relatively constant at approximately $0.46$. As time progresses, the S\&P500 local slope converges to $0.5$ faster than Bitcoin, indicating a quicker convergence to Gaussian diffusion. 
In contrast, the local slope of Bitcoin fluctuates between $0.46$ and $0.5$, gradually converging to normal diffusion over an extended time period. The difference in the rate of convergence to a normal diffusion process between the S\&P500 and Bitcoin (Figure \ref{fig.VOq_alpha}-b) can be attributed to several key factors. Firstly, the high volatility of Bitcoin contributes to its distinct diffusion characteristics. Additionally, the high decentralization of Bitcoin allows it to be influenced by investor sentiment, which further impacts its diffusion dynamics. Lastly, the regulatory framework surrounding cryptocurrency, or rather its absence within traditional regulatory frameworks, adds to the unique behaviour observed in Bitcoin.

A numerical estimation of the value of $1/\tilde{\alpha_t}$ was made using Eq. \ref{Eq.alpha_tilde}, which is shown in Fig.~\ref{fig.VOq_alpha}-c. Using this function, one is able to calculate $G(t,t_0)$ using  and~\ref{Eq.g_s}, and $g(t)$ as its time drivative (Eq.~\ref{Eq.g_t}). Here we applied an approximation in the derivative to the Bitcoin time series to eliminate the negative values. As the final step, we are able to estimate the variable order exponents using the relation~\ref{Eq.g_t}. More precisely, the parameter $D_{t}t^{\xi_{t}-1}$ is calculated for both time series and plotted in Figure \ref{fig.VOq_alpha}-d. Apart from the stochastic fluctuations, we observe that this parameter decreases with time in the small time regime, showing that the diffusion process becomes slower over time as becomes constant for a time period. The decrease of the diffusion coefficient with time is consistent with previous analysis of the S\&P500 data \cite{arias2021methods}. Note that the PDF of the detrended data becomes constant when the time reaches the detrended time window. Thus,  the diffusion coefficient should also become zero when the time reaches the time windows used to detrend the time series. The explanation of this trend of the diffusion coefficient is elusive but there is no reason to expect that the diffusion coefficient remains constant in VO diffusion processes. 



\section{Discussion}

We have proposed a variable-order differential equation to describe nonlinear fractional diffusion processes with anomalous diffusion, applicable for the regime $1\leq q<2$. In the variable-order equation, the analytical solution is self-similar in the broad sense, with $x\sim \phi(t)$ as the scaling variable, which is given in terms of the $q-$Gaussian distribution. In the constant-order equation, the scaling variable reduces to $x\sim t^H$, where $H=1/\alpha$ is the Hurst exponent and $\alpha$ defines whether the diffusion process is either normal or super/sub diffusive.
The variable-order exponents $\alpha(t)$ and $q(t)$ are related to time correlations and non-linear diffusion. The time-dependency of the exponents allows them to comply with the Central Limit Theorem, which requires $\alpha(t)\to 2$ and $q(t)\to 1$ as $t\to\infty$. Typically, alpha is related to the Hurst exponent ($H=1/\alpha$) of the time series associated with the diffusion process. This can be discussed in light of the fractional Brownian motion (fBm), which is a self-similar Gaussian stochastic process characterized by stationary power-law correlated increments. In the fBm, the Hurst exponent provides a crucial link between the diffusion coefficient $\alpha$ through the relation $H=1/\alpha$ \cite{gharari2021space}. This exponent measures the long-range memory in time series data and can also be associated with autocorrelation patterns. Specifically, for $0<H<0.5$ (or $\alpha>2$) the time series exhibits anti-correlated behavior. In the case of $H=0.5$ (or $\alpha=2$), the time series is uncorrelated. Finally, for $0.5<H<1$ (or $\alpha<2$), the time series displays positively correlated behaviour \cite{Kantelhardt2022multi}. Moreover, in positive long-range correlated series, the autocorrelation function follows a power-decay pattern described by $C(s) \sim s^{-\gamma}$ where $\gamma=2-2H$ \cite{Kantelhardt2022multi}.  

The dynamics of the market ecosystem suggest that traders' behavior can influence autocorrelation patterns. In the stock market, traders employ various strategies, including negative trading strategies where they buy after price increases and sell when prices decline, and negative traders who follow the 'buy-low sell-high approach  \cite{koutmos2001positive}. 
Sentana and Wadhwani \cite{Sentana1992feedback} explored the connection between volatility, returns autocorrelation, and trading strategies using a GARCH model. They conducted empirical investigations using the Dow Jones index data and discovered that positive traders can lead to negative autocorrelation, while negative traders can result in positive autocorrelation. Furthermore, they found that in an index comprising numerous securities with different trading frequencies, positive cross-autocorrelation emerges, contributing to positive autocorrelation within the index. This finding aligns with our study, specifically in the context of the S\&P500 market index.

In our analysis, the variable order exponents are observed in both stock markets and cryptocurrencies. In the S\&P500 market $\alpha(t)<2$ while gently increasing to the value  $\alpha(t)=2$ for large times. This behavior is consistent with the ecology of this market, consisting of short to moderate-time investors that lead to short-time correlations. The behavior is different in the Bitcoin index where $\alpha(t)$ oscillates slightly above $2$, indicating anticorrelation in the time series. This behavior is consistent with the peculiar ecology of the cryptocurrencies mainly dominated by speculators who are frequently changing their strategy to maximize utilities. For large times, the time series of price returns become uncorrelated, leading to the expectation that $\alpha$ converges to $2$ in this limit.

The underlining mechanism of the values of the exponent $q(t)$ may be attributed to the interaction between the equities of the stock market, This can be understood from the Langevin equation of the non-linear FPE (Eq.~\ref{eq:f}) that is converted by using the property of the Katugampola derivative $d^\xi/dt^\xi = t^{1-\xi} d/dt$ to \cite{katugampola2014new}

\begin{equation}
\dfrac{\partial P(x,t)}{\partial t}=Dt^{\xi-1} \dfrac{\partial^{2} P(x,t)^{2-q}}{\partial x^2},
\label{eq:fPME}
\end{equation}

The corresponding Langevin equation describing the stochastic dynamics of the price return $X(t)$ has been derived by Bourland \cite{borland1998microscopic}
\begin{equation}
X(t+dt) = X(t)+\eta(t)(Dt^{1-\xi} P(X(t),t)^{1-q})^{1/2}dt,
\label{eq:fSDE}
\end{equation}
where $\eta(t)$ is the white noise signal. Let us consider the case of an idealized gas of particles experiencing Brownian-like motion, For $q=1$ and $\xi=1$ the stochastic dynamic corresponds to the classical random walk that describes ideal gases. For more dense situations, $q$ may take values larger than zero, indicating that the random walk of each particle is influenced by the local density of the particles around its location. This physical picture can be extrapolated to financial markets as follows: Assuming that $X(t)$ is the index of a particular stock, the dependency of the fluctuations on its probability density functions may be attributed to the interaction between different equities in the stock market. In fact, the S\&P500 index is calculated based on the 500 largest companies in the US, and it should be noted that the performance of these companies cannot be assumed to be independent. The exponent $q\approx 1.4$ is observed for price returns calculated within $t=10^3$ minutes (roughly 3 days). For larger times $q$ converges to $2$ since the Central Limit Theorem (CLT) requires the PDF of the price return to converge to a Gaussian distribution when $t\to\infty$. In the case of Bitcoin, the exponent $q$ is larger ($q\approx 1.5$) and it converges to $q=2$ faster (in the order of hours), which may be directly related to the transaction between different crypto-currencies. In both cases, the CLT is guaranteed since the standard deviation of the $q-$Gaussian distribution is finite for $q< 5/3$. Further investigation of these peculiar dynamic features from microscopic models such as agent-based models or order book models would provide some light on the underlines mechanism of the time evolution of these exponents.
\section{Authors' Contribution}
Y.T. and F. G. contributed as first authors of this paper;  F.G. wrote the first draft of the paper.
\appendix
\section{dimensional analysis}~\label{App:dimensional}
In this appendix, we describe the scaling properties of the PDFs. From the dimensional analysis of the normality condition, one realizes that the dimension of any PDF $P(\textbf{x})$ is $[P(\textbf{x})]=[\textbf{x}]^{-d}$, where $[Q]$ shows the \textit{space} dimension of the quantity $Q$. To see this more explicitly, we apply the scaling transformation $\textbf{x}\to \textbf{y}\equiv \lambda \textbf{x}$, and using the conservation of probability $P(r)d^dx=P(y)d^dy$, one finds:
\begin{equation}
P(\lambda x)=\lambda^{-d}P(x).
\label{Eq:scalingPDF}
\end{equation}
For a stochastic process, where $P$ changes with time we use the root mean displacement relation $r\equiv |\textbf{x}|$:
\begin{equation}
\left\langle r^2\right\rangle\propto \phi(t)^2
\end{equation}
to predict the form of the PDF, where $\phi(t)$ is a function of time the form of which determines the anomalous diffusion nature. In this case, the equation~\ref{Eq:scalingPDF} generalizes to:
\begin{equation}
P(\lambda \textbf{x},\lambda\phi(t))=\lambda^{-d}P(\textbf{x},\phi(t)),
\end{equation}
which is realized using the normality condition:
\begin{equation}
\begin{split}
 \int_{-\infty}^{\infty}P&(\textbf{x},\phi(t))d^dx =\lambda^d\int_{-\infty}^{\infty}P(\lambda \textbf{x},\lambda\phi(t))d^dx=1\\
 &\rightarrow P(\lambda \textbf{x},\lambda\phi(t))=\lambda^{-d}P(\textbf{x},\phi(t)).
\end{split}
\end{equation}
A solution of this equation is a factorized form as follows:
\begin{equation}
P(\textbf{x},t)=\frac{1}{ \phi(t)^d}  F  \left[\phi(t)^{-1}\textbf{x} \right],
\label{Eq:generalFormApendix}
\end{equation} 
where the function $F$ is a well-behaved one to be determined using the governing equation. Note that $F$ is invariant under $\textbf{x}\to \lambda\textbf{x}, \phi(t)\to\lambda\phi(t)$, and also:
\begin{equation}
    \left\langle r^2\right\rangle=\frac{\int \text{d}^dxr^2F\left[\frac{\textbf{x}}{\phi(t)} \right]}{\int \text{d}^dxF\left[\frac{\textbf{x}}{\phi(t)} \right]}=\left[\frac{\int \text{d}^duu^2F\left[\textbf{z} \right]}{\int \text{d}^duF\left[\textbf{z} \right]}\right] \phi(t)^2,
    \label{Eq:scalingAPP}
\end{equation}
where $\textbf{z}\equiv \frac{\textbf{x}}{\phi(t)}$. The scaling properties of the time series are associated with the form of $\phi(t)$. In fact, for the solution of the Eq.~\ref{eq:f} we have $\phi(t)=\phi_{\text{SS}}(t)$ where the index ``SS'' points out the self-similarity law, given by~\cite{gharari2021space,alonso2019q}:
\begin{equation}
\begin{split}
\phi_{\text{SS}}(t)\propto t^{1/\alpha}\equiv t^{-H}.
\end{split}
\label{eq:m}
\end{equation}
where $\alpha$ is a self-similarity exponent, and $H$ is the Hurst exponent.
Combining Eq.~\ref{Eq:scaling} and Eq.~\ref{eq:m}, one reaches the Eq.~\ref{Eq:scaling0}.
\section{Relationship between FBM, time fractional FPEs  and  SDEs }\label{App:FBM}
 This short appendix is devoted to some scaling properties of the fractional Brownian Motion (fBm), especially by focusing on its Langevin equation, and the corresponding Fokker-Planck equation (FPE). Our primary objective is to explore the relationship between time-fractional generalized Lagevine and FPE equations for fBm. This example sheds light on the essential difference between the driving processes of BM as an example of semi-martingales and fBM as a representative of non-semi-martingales. The complexity of the latter case arises mainly from the presence of correlations, resulting in sub- or super-diffusion. This also shows the consequences of the presence of non-local effects~\cite{gharari2021space}.

The classical FPE establishes a relationship between the SDE driven by Bm and its associated partial differential equation (PDE). The fBm  $B^H:=\lbrace B^H(t), t\geqslant 0\rbrace$
is a family of stochastic processes indexed by the Hurst index $H \in (0, 1).$ For each $H$, the process $B^H$  is defined as a weighted moving average of a Bm process:
\begin{equation}
    B^H(t)=\int_0^t{(t-\tau)^{H-1/2}dB(\tau)},
\end{equation}
where $B:=\lbrace B(t), t\geqslant 0\rbrace$ is the Brownian motion and $H$ is the Hurst parameter. The fBm with variance $\sigma(t)$ is the solution of the following fractional SDE \cite{di2022fokker}:
\begin{equation}
    \frac{d^\xi X}{d^\xi t}=\sigma(t)\eta(t)
\end{equation}
where $\eta(t)=dB/dt$ is the normalized white noise, $ 1/2 < \xi=H+1/2 \leq 1.5$  and  $d^\xi/d t^\xi$ is the Riemann–Liouville fractional operator that is defined as
\begin{equation}
    \frac{d^\xi f}{dt^\xi}:=\frac{1}{\Gamma(\xi)}\int_0^t{(t-\tau)^{\xi-1} f(\tau) d\tau}
\end{equation}
Further, the fractional FPE of the fBm process is given by \cite{di2022fokker}
\begin{equation}
    \frac{d^{2H}P_X(x,t)}{dt^{2H}}=\frac{\Gamma(2H)}{2\Gamma^2(H+1/2)}\sigma(t)\frac{\partial^2 P_X(x,t)}{\partial x^2}
\end{equation}
There are two main differences between the fBm and the fractional q-Gaussian process investigated in this paper: first, the FPE of the fBm is linear and its solution is expressed in terms of Gaussian distribution, while the latter is nonlinear and the solution is given in terms of the q-Gaussian distribution. Second, the fractional FPE of the fBm involves non-local fractional derivatives while the fractional q-Gaussian process involves local (Katugampola) fractional operators. Yet in both cases, one can recover the Bm by either taking $H=1/2$ in the fBm or $q=1$ and $\alpha=1$ in the fractional q-Gaussian diffusion.

On the other hand, discussing the relationship between the fractional
FPE (or the governing equation) and the SDE of their associated time series are equivalent to discussing the relation between the auto-correlation of the time series and the fractionalization scheme in the governing equation. The key point is the scaling properties of the governing
equation describing a self-similar time series. If $\lbrace X (t)\rbrace_{t\in \mathbb{Z}}  $  denotes a self-similar time series with the
property
\begin{equation}
 \sqrt{\left\langle X^{2}\right\rangle} \propto t^{H}  
\end{equation}
Then the corresponding governing equation should have the same symmetry. It means the invariance of the PDF (of the governing equation) up to a scaling factor, i.e. 
\begin{equation}\label{FPE_FBM1}
 P(X,t)=t^{-H}F(X/t^{H}),
 \end{equation}
  where $F$ is a function to be fixed by the governing equation. 
FBMs are described by Eq. (\ref{FPE_FBM1}) when $F$ is an exponential function,
\begin{equation}\label{FPE_FBM2}
 P\left(X,t \right)\propto t^{-H}\exp \left[-\frac{1}{2}\left(\frac{X}{t^H} \right)^2 \right].
 \end{equation}
In fact, the governing equation for a self-similar time series should include fractional operators, or one should use a space- or time-dependent diffusion coefficient with power-law dependence. The symmetry of the
system can easily be found in the corresponding PDF of
the time series, and also by calculating the second moment of $X,$ which $X$ denotes a self-similar time series. This means the fractionalization exponent is manifest in the PDF.\\
Consider a general $H$-self-similar time series $\lbrace X (t)\rbrace_{t\in \mathbb{Z}}  $ defined by the relation  $\lbrace X (ct)\rbrace_{t\in \mathbb{Z}} \equiv \lbrace c^{H} X(t)\rbrace_{t\in \mathbb{Z} }, $   where $c>0,$   and $ H$ is the Hurst exponent. If this process has stationary increments $ Y_{n}=X (n)-X (n-1),$ then the auto-correlation $\gamma_{Y}(k)\equiv \langle Y_{k} Y_{0} \rangle- \langle Y_{k} \rangle\langle  Y_{0} \rangle$ behaves like $ k^{ 2d-1}$  as $ k \rightarrow \infty,$ where $d=H-\frac{1}{2}$, and $ 0 < d < 1/2$, ensuring that
$\sum_{k=-\infty}^{\infty}\gamma_{Y}(k)=\infty$.
  From a spectral domain perspective, the
spectral density of $ \lbrace Y_{n}\rbrace$ behaves as $\omega^{-2d}$
  as the frequency $\omega \rightarrow 0$. This relates self-similar time series with fractionalization which is applied to the price return time series which becomes stationary by normalizing the detrended data set (see Eq.~\ref{IV_EQ1})~\cite{gharari2021space}. 
 Let us consider the  special case  
$\Phi(t) \sim t^{1/\alpha_t}$ in Eq. \ref{eq:second_moment}. 
 Following these facts, we suggest a relation between the Hurst exponent $H_{t}=1/ \alpha_{t} $ given in Eq.~(\ref{eq:d_e_x}), with a self-similar function Eq.~\ref{Eq:PDF}  for the VO-nonlinear case.

\section{Some details of VO calculations for PME}
\label{App:Katugampola-derivative}
In this appendix, we introduce and inspect a VO extension of the Katugampola fractional operator, denoted as VO-K in this paper. We outline the definition of this operator and discuss some of its key properties. For a more comprehensive understanding of the Katugampola fractional operator and its underlying principles, we recommend referring to~\cite{katugampola2014new}.

A VO-K fractional operator, which is used to construct the VO-PME in Sec.~\ref{SEC:1} is defined as
\begin{equation}
\mathcal{D} ^{\alpha(t)}f( t)=\lim_{\epsilon \to 0}\dfrac{f(te^{\epsilon t^{-\alpha(t)}})-f(t)}{\epsilon},
\label{Eq:Katugampola}
\end{equation}
for $ t>0  $ and $ \alpha(t)\in(0,1]$.
If $0 \leq \alpha(t)<1,$ the VO-K operator generalizes the
classical calculus properties of polynomials. Furthermore, if $\alpha(t)=1,$ the definition is equivalent to the
classical definition of the first-order derivative of the function $f$.  
When $\alpha(t) \in (n, n+1]$ (for some $n \in \mathbb{N}$, and $f$ is an $n-$differentiable at $ t>0$), the above definition generalizes to
\begin{equation*}\label{ufo8}
\mathcal{D} ^{\alpha(t)}f(t)=\lim_{\epsilon \to 0}\dfrac{f^{(n)}(t e^{\epsilon x^{n-\alpha(t)}})-f^{(n)}(t)}{\epsilon}.
\end{equation*}
If $f$ is $(n+1)-$differentiable at $t>0, $ then we have
\begin{equation}\label{c2_app}
\mathcal{D} ^{\alpha(t)}f(t)=t^{n+1-\alpha(t)}f^{(n+1)}(t).
\end{equation}
The properties of the VO-K derivatives are a simple extension of the ordinary derivatives:
 \begin{align*}
 \mathcal{D} ^{\alpha(t)}[af+bg] &=a\mathcal{D}^{\alpha(t)}(f)+b\mathcal{D} ^{\alpha(t)}(g), \,\,\,\forall a,b\in\mathbb{R},\\
 \mathcal{D} ^{\alpha(t)}[C]&=0, \,\,\,C\in\mathbb{R},\\
 \mathcal{D} ^{\alpha(t)}[fg]&= f\mathcal{D} ^{\alpha(t)}(g)+g\mathcal{D} ^{\alpha(t)}(f),\\
 \mathcal{D} ^{\alpha(t)}[f/g]&=\frac{g\mathcal{D} ^{\alpha(t)}(f)-f\mathcal{D} ^{\alpha(t)}(g)}{g^{2}},\\
  \mathcal{D} ^{\alpha(t)}(f\text{o}g)(t)&= f^{ \prime}(g(t))\mathcal{D} ^{\alpha(t)}g(t),
\end{align*}  
where $f\text{o}g(t)\equiv f(g(t))$. The proof of the properties of the variable order Katugampola fractional operator (VO-K) is analogous to the proof presented in \cite{katugampola2014new}. In our case, we replace the constant order parameter $\alpha$ with the variable order parameter $\alpha(t)$.
%

  Throughout this paper, the notations $\mathcal{D}^{\alpha(t)}$ and $\frac{\partial^{\alpha(t)}}{\partial t^{\alpha(t)}}$ is used with the same meaning. The fractional Katugampola calculations for the VO-PME is performed, as done in SEC.~(\ref{SEC:1}),
  starting from Eq.~\ref{Eq:generalForm}, and the Eq.~(\ref{eq:d_e_x}), and applying the properties of the VO-K derivative, we get
  \begin{align}\nonumber
 \frac{\partial^{\xi(t)}}{\partial t^{\xi(t)}}& \left(  \frac{1}{\phi(t)} F \left(\frac{x}{\phi(t)} \right) \right)  
  =F \left(\frac{x}{\phi(t)} \right) \left( \dfrac{-\mathcal{D}^{\xi(t)}\phi(t)}{\phi^{2}(t)}  \right)\\\nonumber
 &+\frac{1}{\phi(t)} \left(  F^{^{\prime}}\left(\frac{x}{\phi(t)}\right)\mathcal{D}^{\xi(t)} \left(\frac{x}{\phi(t)} \right) \right)\\\nonumber
&=F\left(z\right) \left( \dfrac{-\mathcal{D}^{\xi(t)}\phi(t)}{\phi^{2}(t)} \right)+ z  F^{^{\prime}}\left(z\right) \left(\dfrac{-\mathcal{D}^{\xi(t)}\phi(t) }{\phi^{2}(t)} \right)\\
&= \dfrac{-\mathcal{D}^{\xi(t)}\phi(t) }{\phi^{2}(t)} \frac{d}{dz} [zF(z)]. 
\end{align}
Then, we get the following two equations:   
\begin{equation}
\begin{split}
&\partial^{2}_{x}P^{\nu(t)}(x,t) = \frac{1}{\phi^{\nu(t)+2}}\frac{d^{2}}{dz^{2}}F^{\nu(t)},\\
&\dfrac{\partial^{\xi(t)}P(x,t)}{\partial t^{\xi(t)}}=\dfrac{-1}{\phi^{2}(t)}\dfrac{\partial^{\xi(t)} \phi}{\partial t^{\xi(t)}}\left[F+z\frac{d}{dz}F\right],
\end{split}
\end{equation}
so that,
\begin{equation}\label{equ.d3}
\dfrac{-1}{\phi^{2}(t)}\dfrac{\partial^{\xi(t)}\phi}{\partial t^{\xi(t)}}\frac{d}{dz}[ zF]=\dfrac{ D (t)}{\phi^{\nu(t)+2}}\frac{d^{2}}{dz^{2}}F^{\nu(t)}.
\end{equation}\label{equ.D3}

\section{The local VO fractional non-linear time diffusion equation with drift}~\label{SEC:drift}

The drift is often an inevitable part of stochastic systems, that should be analyzed in detail for every case study to control its effects. Although it is suggested to define the equations for the general drift term. For the case where it depends only on time (as is the case for many physical systems of interest), the situation becomes easier. In this case, the governing equation is:
\begin{equation}\label{general eq}
\begin{split}
\dfrac{\partial^{\xi(t)}}{\partial t^{\xi(t)}}&P(x,t)=-a(t)\dfrac{\partial P (x,t)}{\partial x}+  D (t)\dfrac{\partial^{2} P^{\nu(t)} (x,t)}{\partial x^{2}}, 
\end{split}
 \,\,\,\,\,  
\end{equation}
Through a change of variable $\tau= t^{\xi(t)}$ and VO Katugampola (VO-K) fractional derivative (see Appendix~\ref{App:Katugampola-derivative}, provided $ h(t)=\xi(t) \ln t $ has an inverse function, we have:
\begin{equation*}
\partial_{\tau} P(x,t) =- a_{1} (\tau)\partial_{x}P(x,t)+ D _{1}(\tau)\partial_{x}^{2}P^{\nu_{1}(\tau)}(x,t),
\end{equation*}
where $ a_{1} (\tau)=a(t(\tau))(t \xi^{\prime}(t)\ln t+\xi(t))^{-1},\,D_{1}(\tau)= D (t(\tau))(t \xi^{\prime}(t)\ln t+\xi(t))^{-1},$ and $\nu_{1}(\tau)=\nu (t(\tau))$. By using the change of variable $(s,y)= (\tau,x-x_{0}-f(\tau))$, where $f(\tau)=\int_{0}^{\tau} a_{1} (\tau^{\prime} )d \tau^{\prime} $, and using the fact that $\frac{\partial y }{\partial \tau }=-a_{1}(\tau)$ and $\partial_{\tau}+a_{1}(\tau)\partial_{x}=\partial_{s}$, one finds that  the governing equation  $P(y,\tau)$ is:
\begin{equation*}
\partial_{\tau}P(y,\tau)= D (\tau)\partial_{y}^{2}P^{\nu (\tau)}(y,\tau),
\end{equation*}
for which the solution is ($x_{0}\equiv 0$ and $k\equiv 1$ and $k_1\equiv 0$): 
\begin{align}\nonumber 
P&(y,\tau\vert y_{0},\tau_{0})=\frac{A _{q}(t)} { \left( \int_{\tau_{0}}^{\tau}g(s)ds
\right)^{\frac{1}{\nu (\tau)+1}}}\\ &\times\left(c +\dfrac{(\nu(\tau)-1) }{2\nu(\tau)}\frac{y^{2}}{ \left(\int_{\tau_{0}}^{\tau}g(s)ds\right)^{\frac{2}{\nu(\tau)+1}}} \right)^{\frac{1}{\nu(\tau)-1}},
\end{align}
where $g(s)= D (s)(1+\nu (s)).$
 Let us equate  the $P(y,\tau)$: 
\begin{equation*}
P(x,t)=\frac{\partial{y}}{\partial {x}}P(y,\tau(t)).
\end{equation*}
 Then, we obtain that:
\begin{equation}
\begin{split}
P&(x,t\vert_{0},t_{0})=\frac{A_{q}(t)}
{\left(\int_{t_{0}}^{t}g(s)ds\right)^{\frac{1}{\nu(t)+1}}}\\ 
&\times\left(c+\dfrac{(\nu(t)-1) }{2\nu(t)}\frac{(x-f(t))^{2}}{\left( \int_{t_{0}}^{t}g(s)ds\right)^{\frac{2}{\nu(t)+1}}}\right)^{\frac{1}{\nu(t)-1}},
\label{Eq:LPME_drift}
\end{split}
\end{equation}
where $A_{q}$ is a normalization factor, and $c$ and $k$ are constant. The Eq.~(\ref{Eq:LPME_drift}) is a vo$q$-Gaussian solution with a drift.


\bibliography{refs}
\end{document}